\documentclass[pra,twocolumn,showpacs]{revtex4}
\usepackage{epsfig}
\usepackage{graphicx}
\usepackage{dcolumn}
\usepackage{amsmath,amssymb}
\usepackage{pstricks}
\usepackage{color}
\usepackage{footnote}
\usepackage{bm}
\usepackage{bbold}
\usepackage[colorlinks=true,citecolor={blue},urlcolor={blue}, linkcolor=blue]{hyperref}
\begin{document}
\title{Role of electron-correlation in the $\mathcal{P,T}$-odd effects of CdH: A relativistic coupled-cluster investigation} 
\author{Kaushik Talukdar,$^{1,}$\footnote{talukdar.kaushik7970@gmail.com}
Malaya K. Nayak,$^{2,}$\footnote{mknayak@barc.gov.in, mk.nayak72@gmail.com}
Nayana Vaval,$^{3,}$\footnote{np.vaval@gmail.com}
Sourav Pal$^{4,1,}$\footnote{spal@chem.iitb.ac.in}}
\affiliation{$^1$Department of Chemistry, Indian Institute of Technology Bombay, Powai, Mumbai 400076,  India}
\affiliation{$^2$Theoretical Chemistry Section, Bhabha Atomic Research Centre, Trombay, Mumbai 400085, India}
\affiliation{$^3$Electronic Structure Theory Group, Physical Chemistry Division, CSIR-National Chemical Laboratory, Pune 411008, India}
\affiliation{$^4$Department of Chemistry, Indian Institute of Science Education and Research Kolkata, Mohanpur 741246, India}
\begin{abstract}
We investigate the parity ($\mathcal{P}$) and time-reversal ($\mathcal{T}$) symmetry violating effects in the CdH molecule and perform the relativistic coupled-cluster
calculation of the molecular parameters- $E_\text{eff}$, $W_\text{s}$ and $W_\text{M}$ related to the electric dipole moment of electron (eEDM) interaction,
the scalar-pseudoscalar (S-PS) nucleus-electron neutral current coupling and the nuclear magnetic quadrupole moment (MQM) interaction with electrons, respectively.
We also compute the molecular dipole moment and the magnetic hyperfine structure coupling constant of CdH. The value of $E_\text{eff}$, $W_\text{s}$ and $W_\text{M}$ 
obtained by us in the said molecule are 12.2 GV/cm, 14.0 kHz 
and $0.82\times$10$^{33}$ Hz/e cm$^2$, respectively, with an uncertainty of 10\%.
Furthermore, we study the trend of electron-correlation in the computed properties of CdH and that 
of the $\mathcal{P,T}$-odd parameters in the group-12 monohydrides (i.e., ZnH, CdH, and HgH).

\end{abstract}
\maketitle
\section{Introduction}

The well-known particles and forces are not able to explain the universe almost entirely comprised of matters and thus, there is a strong
possibility of the existence of ``new particles and forces'' (which are unknown to date). 
The so-called ``new physics'' beyond the standard model (SM) of elementary particles has been emerging as a bright area of research to search the so-called ``new particles and forces''. 
Violations of charge conjugation ($\mathcal{C}$) and parity ($\mathcal{P}$) or time-reversal ($\mathcal{T}$) invariance beyond the standard 
model can explore this ``new physics'', which in turn, helps to unravel the mystery of matter-antimatter asymmetry of our universe. 
The phenomenon of $\mathcal{CP}$ violation results the
intrinsic electric dipole moment of the electron (eEDM ($d_e$)) \cite{bernreuther_1991, tl_edm, ybf_edm, titov_tho, tho_edm}, 
the scalar-pseudoscalar (S-PS) nucleus-electron neutral current coupling \cite{tho_edm, sasmal_raf, kudashov_2014, sudip_hgh} and the nuclear magnetic quadrupole 
moment (MQM) \cite{fleig_tan, flambaum_2017, titov_2014}. 
According to the SM, the $d_e$ is so small ($< 10^{-38}$ e.cm \cite{khriplovich_2011}) that it can not
be experimentally observed. 
But many extensions of the SM predict the eEDM to be in the range of $10^{-29}-10^{-26}$ e.cm \cite{commins_1999} and
the sensitivity of the modern eEDM experiment is also found to be in the same range.
The best upper bound limit of eEDM ($< 1.3 \times 10^{-29}$ e.cm) is recently obtained in the ThO-experiment carried out by the ACME collaboration \cite{tho_new}.
It is well-known that the eEDM effect is strongly enhanced in heavy polar diatomic paramagnetic molecules due to their high internal effective electric 
field ($E_\mathrm{eff}$). 
In addition to the eEDM, the S-PS neutral current coupling and the MQM-electron interaction 
are the other main possible sources of the permanent electric dipole moment (EDM) in paramagnetic systems. In the $\mathcal{P,T}$-odd frequency shift experiment,
the permanent EDM of the experimental candidate interacts with the electromagnetic field resulting in a shift in energy (i.e., frequency) that can be measured.
To analyse this frequency shift in terms of the eEDM, the fundamental S-PS coupling constant ($k_s$) and the magnetic quadrupole moment ($M$) of the nucleus,
the accurate values of $E_\mathrm{eff}$, scalar-pseudoscalar $\mathcal{P,T}$-odd interaction parameter ($W_\mathrm{s}$), and 
the MQM interaction constant ($W_\mathrm{M}$), respectively are required. Moreover, a large value of a molecular $\mathcal{P,T}$-odd interaction parameter implies that
the corresponding $\mathcal{P,T}$-odd interaction in that molecule may be significantly large. But
the value of $E_\mathrm{eff}$, $W_\mathrm{s}$ and $W_\mathrm{M}$ 
cannot be experimentally measured and can only be calculated using the highly accurate electronic structure theories.
The $\mathcal{P,T}$-odd molecular parameters are also known as the ``atom-in-compound'' (AIC) \cite{titove_aic} properties because the operators corresponding to these
properties are heavily concentrated on nuclei or in atomic cores. Usually, the AIC properties strongly depend on the electronic configuration of a particular
atom in a compound rather than on the chemical bond between atoms. The magnetic hyperfine structure (HFS) interaction constant is another important AIC property.
These properties are very sensitive to the valence electron density (i.e., the wave function) near the nuclear region of the heavy atom and  
therefore, can be accurately calculated using an {\it ab initio} method that can efficiently incorporate both the relativistic and electron-correlation effects.

In the single-reference framework, although the Dirac-Hartree-Fock (DHF) method
can treat the relativistic motion of electrons, it misses the correlation effects of electrons in an atom or a molecule. Therefore, the post-DHF 
methods are necessary to incorporate the correlation effects of electrons. The many-body methods such as the configuration interaction (CI) \cite{szabo_ostlund}, the Moller-Plesset perturbation
theory (MBPT) \cite{szabo_ostlund} and the coupled-cluster (CC) \cite{szabo_ostlund,cizek_1966,cizek_1967,bartlett_1978} etc. are widely used in the literature for the treatment of 
electron-correlation. It is worth  mentioning that the multi-reference many-body 
theories are used to incorporate the static correlation effects. Nonetheless, for the efficient treatment of the dynamic electron correlation in many-electron
systems the single-reference coupled-cluster (SRCC) method has been considered as one of the most suitable tools. Moreover, the properties of atoms and molecules can be calculated
using either the energy-derivative method or the expectation-value approach within the SRCC framework. The SRCC method, usually being a nonvariational approach, does not satisfy
the generalized Hellman-Feynman theorem. Also, the energy-derivative method and the expectation-value approach are not same in nonvariational 
framework \cite{monkhorst_1977, bartlett_1984}. 
The energy-derivative in the nonvariational model contains the corresponding expectation-value plus some additional terms, which leads to the fact that the property obtained by
the energy-derivative technique is closer to that evaluated by the full CI method than the property value calculated using the expectation-value approach \cite{zvector_1989}.
We have already mentioned that both the relativistic and electron-correlation effects are the keys for the precise calculation of the AIC properties.
Thus, for this purpose, the relativistic single-reference coupled-cluster method could be a suitable tool as it can deal with both the effects of correlation and
relativistic motion of electrons. The $Z$-vector method \cite{schafer_1984, zvector_1989} is a popular energy derivative approach to calculate 
the first-order properties of atoms and molecules. In recent times, Sasmal {\it et al.} \cite{sasmal_pra_rapid} introduced the $Z$-vector technique into the four-component relativistic 
coupled-cluster realm and successfully employed the method to calculate various AIC properties of atoms, ions and molecules. It is found that
the $Z$-vector method can produce precise results of the AIC properties in the relativistic SRCC framework. 

As mentioned above the $\mathcal{P,T}$-odd interaction parameters in polar heavy molecules are usually high in magnitude and thus, these molecules are always expected to be good
candidates for the $\mathcal{P,T}$-odd frequency shift experiments. However, the large value of $\mathcal{P,T}$-odd interaction parameter in a molecule is not sufficient for
the success of such an experiment. The experimental molecule must be easily polarizable to fully utilize the applied electric field.
Usually, a molecule with small rotational constant and large dipole moment is easily polarizable in a spectroscopic experiment. 
But recently, Kozlov {\it et al.} \cite{kozlov_hgh} suggested an alternative route to carry out the eEDM
experiment using heavy diatomic radicals. They showed that the less polar molecule such as HgH can be polarized easily in the matrix isolated nonspectroscopic solid-state 
experiment to measure the eEDM. Interestingly, the small dipole moment of the diatomic molecular radical is one of the important conditions to achieve
higher sensitivity for the eEDM in
the said non-spectroscopic solid-state experiment. The internuclear distance of the HgH molecule is 1.7\AA. It can be easily trapped in the
Ar-matrix of the cell size 4.5 \AA. 
CdH is a molecule having similar characteristics as that of HgH. 
The dipole moment and internuclear distance of CdH is close to that of HgH. Thus, CdH can also be expected as a possible candidate for the non-spectroscopic solid-state experiment 
and hence, theoretical study of CdH in search of the $\mathcal{P,T}$-odd effects could be important. 
Recently, Berger and co-workers {\red \cite{berger_2019}} performed a systematic study of the relativistic and chemical enhancements of the 
$\mathcal{P,T}$-odd properties in various diatomic radicals including ZnH, CdH, and HgH using a quasirelativistic approach within the framework of 
complex generalized Hartree-Fock (cGHF) or Kohn-Sham (cGKS). In that work, the periodic trend of the $\mathcal{P,T}$-odd effects was discussed in detail. 
However, it is worth mentioning that the calculations of $\mathcal{P,T}$-odd properties of molecules are often very difficult and challenging due to the strong correlations 
between the electrons. That is why a systematic study of the role of electron-correlation effects and its trend in the calculations of these properties is extremely important.
Use of a more robust method, precisely speaking, a fully relativistic coupled-cluster method would be helpful for a better understanding of the said effects in the molecules.
Therefore, in this work, we have calculated the $E_\text{eff}$, $W_\text{s}$ and $W_\text{M}$ of CdH in its ground electronic
($^2\Sigma_{1/2}$) state and studied the correlation trend in these properties using the $Z$-vector method in the domain of four-component
relativistic coupled-cluster theory. The magnetic HFS constants of CdH are also calculated to estimate the accuracy of the employed
method. We also compute the $\mathcal{P,T}$-odd parameters of ZnH and HgH to see the trend of the calculated $\mathcal{P,T}$-odd molecular parameters in the group-12 monohydrides.

The structure of the paper is as follows. The important aspects of the theory of the calculated properties and those of
the $Z$-vector approach in the domain of relativistic SRCC method are discussed in Sec. \ref{theory}.
Computational details are given in Sec. \ref{comp}. The results of the present work are presented and discussed 
in Sec. \ref{res_dis}. Finally, the conclusion of the present study is given in Sec. \ref{conc}. 
Atomic units are used explicitly in this article unless stated. 

\section{Theory}\label{theory}
\subsection{One-electron property operators}\label{prop}
The internal electric field ($E_{\text{eff}}$) experienced by the unpaired electron can be defined by the following matrix element:
\begin{eqnarray}
 E_{\text{eff}} = |W_d \Omega|  = | \langle \Psi_{\Omega} | \sum_j^n \frac{H_d(j)}{d_e} | \Psi_{\Omega} \rangle |,
 \label{E_eff}
\end{eqnarray}
where, $W_d$ is the $\mathcal{P,T}$-odd constant for eEDM interaction, $\Omega$ is the projection of total angular momentum on the internuclear axis ($z$ axis)
of the molecule, $\Psi_{\Omega}$ is the wave function of the $\Omega$ state, and $n$ is the total number of electrons. The value of $\Omega$ is 1/2 for the ground 
electronic ($^{2}\Sigma_{1/2}$) state of CdH, ZnH, and HgH. And, the H$_d$ in the above expression is the Hamiltonian for the
interaction of the eEDM ($d_e$) with the molecular electric field \cite{kozlov_1987, titov_2006}, which is given by
\begin{eqnarray}
 H_d = 2icd_e \gamma^0 \gamma^5 {\bf \it p}^2 ,
\label{H_d}
\end{eqnarray}
where, $c$ is the speed of light, $\gamma$ are Dirac matrices, and {\bf \it p} is the momentum operator.

\par

The S-PS interaction constant, $W_{\text{s}}$ can be evaluated from the following matrix element:
\begin{eqnarray}
 W_{\text{s}}=|\frac{1}{\Omega k_\text{s}}\langle \Psi_{\Omega}|\sum_j^n H_{\text{SP}}(j)| \Psi_{\Omega} \rangle|,
\label{W_s}
\end{eqnarray}
where, $k_s$ is known as the dimensionless nucleus-electron scalar-pseudoscalar coupling constant. This constant is 
defined as Z$k_s$=(Z$k_{s,p}$+N$k_{s,n}$), where Z and N are the number of protons and neutrons, repectively. And, $k_{s,p}$ and $k_{s,n}$
are known as the electron-proton and electron-neutron coupling constant, respectively. $H_{\text{SP}}$ is the interaction Hamiltonian for scalar-pseudoscalar(S-PS) 
nucleus-electron coupling \cite{hunter_1991}, which is defined as follows:
\begin{eqnarray}
H_{\text{SP}}= i\frac{G_{F}}{\sqrt{2}}Zk_{s} \gamma^0 \gamma^5 \rho_N(r) ,
\label{H_SP}
\end{eqnarray}
where, G$_F$ is the Fermi constant, Z is the nuclear charge (i.e., number of protons) and $\rho_N(r)$ is known as the nuclear charge density normalized to unity.

The ratio of $E_{\mathrm{eff}}$ to $W_{\mathrm{s}}$ is known as $R$ \cite{dzuba_2011}, which is very important to set the model independent limit of eEDM and fundamental 
S-PS nucleus-electron coupling constant. It is worth mentioning here that the $R$ has a fixed value for a particular nucleus irrespective of the 
diatom \cite{dzuba_2011}. Using $R$ we can write the relation of
independent $d_e$ and $k_s$ with experimentally determined $d_e^{expt}$ as follows (for more details see Ref. \cite{sudip_hgh}): 
\begin{eqnarray}
 d_e + \frac{k_s}{2R} = d_e^{expt}|_{\!_{k_s=0}}.
\label{relation}
\end{eqnarray}
Here $d_e^{expt}|_{\!_{k_s=0}}$ is the eEDM limit obtained from the $\mathcal{P,T}$-odd frequency shift experiment at the limit $k_s$ = 0.

The Hamiltonian for the interaction of nuclear MQM with the magnetic field produced by electrons \cite{flambaum_2017, kozlov_1987} is given by 
\begin{eqnarray}
H_{\text{MQM}}=-\frac{M}{2I(2I-1)}T_{ik}\frac{3}{2}\frac{[\vec{\alpha} \times \vec{r}]_i r_k}{r^5},
\label{H_MQM}
\end{eqnarray}
where, {\boldmath $M$} is known as the nuclear magnetic quadrupole moment with components
\begin{eqnarray}
M_{ik}=\frac{3M}{2I(2I-1)}T_{ik}, \\
T_{ik}=I_i I_k + I_k I_i-\frac{2}{3}\delta_{ik}I(I+1).
\end{eqnarray}
However, as shown in Ref. \cite{flambaum_1984}, for the subspace of $\pm \Omega$, the Eq. (\ref{H_MQM}) reduces to
\begin{eqnarray}
H_{\text{MQM}}=-\frac{W_M M}{2I(2I-1)}\vec{S}^{\prime}\hat{T}\vec{n},
\end{eqnarray}
where, $\vec{n}$ and $\vec{S}^{\prime}$ are the unit vector along the molecular axis and the effective electron spin, repectively. The 
$W_M$ in the above expression is known as the nuclear MQM interaction constant and is defined by the following matrix element:
\begin{eqnarray}
W_M=|\frac{3}{2 \Omega} \langle \Psi_{\Omega} | \sum_i^n \left( 
       \frac{\vec{\alpha}_i \times \vec{r}_i}{r_i^5} \right)_z r_z | \Psi_{\Omega} \rangle|
\label{W_M}
\end{eqnarray}


The accuracy of the wave function used for the calculations of Eq. \ref{E_eff}, \ref{W_s} and \ref{W_M} can be estimated by comparing the theoretically calculated
HFS interaction constant with the available experimental value, because the HFS constant also depends on a precise wave function near the nuclear region. 
The parallel ($A_{\|}$) and perpendicular ($A_{\perp}$) components of the magnetic hyperfine structure constant
of a molecule can be defined by the following matrix element:
\begin{eqnarray}
A_{\|(\perp)}= \frac{\vec{\mu_k}}{I\Omega} \cdot \langle \Psi_{\Omega} | \sum_i^n
\left( \frac{\vec{\alpha}_i \times \vec{r}_i}{r_i^3} \right)_{z(x/y)} | \Psi_{\Omega(-\Omega)}  \rangle,
\label{hfs_mol}
\end{eqnarray}
where, $\vec{\mu}_k$ is nothing but the magnetic moment of the nucleus $k$.

\par

\subsection{$Z$-vector method in relativistic coupled-cluster singles and doubles framework}\label{corr}
The SRCC wave function has an exponential form and is given as
\begin{eqnarray}
|\Psi_{cc}\rangle=e^{T}|\Phi_0\rangle ,
\end{eqnarray}
where $\Phi_0$ is the Dirac-Hartree-Fock (DHF) determinant and
$T$ is known as the coupled-cluster excitation operator. $T$ is defined as
\begin{eqnarray}
 T=T_1+T_2+\dots +T_N=\sum_n^N T_n ,
\end{eqnarray}
with
\begin{eqnarray}
 T_m= \frac{1}{(m!)^2} \sum_{ij\dots ab \dots} t_{ij \dots}^{ab \dots}{a_a^{\dagger}a_b^{\dagger} \dots a_j a_i} ,
\end{eqnarray}
where $i,j..(a,b..)$ indices are the occupied (unoccupied) spinors and $t_{ij..}^{ab..}$ is the cluster amplitude corresponding 
to $T_m$. In coupled-cluster model with single and double excitation (CCSD), $T=T_1+T_2$, and the unknown cluster amplitudes corresponding to $T_1$ and $T_2$ 
can be obtained by solving the following equations:
\begin{eqnarray}
 \langle \Phi_{i}^{a} | (H_Ne^T)_c | \Phi_0 \rangle = 0 , \,\,
  \langle \Phi_{ij}^{ab} | (H_Ne^T)_c | \Phi_0 \rangle = 0 ,
 \label{cc_amplitudes}
\end{eqnarray}
where, $H_N$ is the normal ordered Dirac-Coulomb (DC) Hamiltonian. The subscript $c$ represents connectedness that 
ensures the size-extensivity. Connectedness means that only the connected terms survive in the contraction between $H_N$ and $T$. 
The DC Hamiltonian is defined as 
\begin{eqnarray}
{H_{DC}} &=&\sum_{j} \Big [-ic (\vec {\alpha}\cdot \vec {\nabla})_j + (\beta -{\mathbb{1}_4}) c^{2} + V^{nuc}(r_j)+ \nonumber\\
       && \sum_{k>j} \frac{1}{r_{jk}} {\mathbb{1}_4}\Big].
\end{eqnarray}
Here, {\bf$\alpha$} and $\beta$ are the conventional Dirac matrices. Furthermore,
${\mathbb{1}_4}$ is the 4$\times$4 identity matrix, $j$ reperesents the electron and
$V^{nuc}(r_j)$ is the potential function for finite size nucleus, defined in terms of a Gaussian charge distribution.

Now, the correlation energy is obtained from the following equation:
\begin{eqnarray}
E_{corr} = \langle \Phi_0 | (H_Ne^T)_c | \Phi_0 \rangle.
\end{eqnarray}

\par
The properties of many-electron atoms and molecules can be obtained by energy-derivative approach within the SRCC framework.
The $Z$-vector method \cite{zvector_1989} is a widely used energy-derivative approach, which has been recently extended 
into the relativistic coupled-cluster domain by Sasmal {\it et al.} \cite{sasmal_pra_rapid}. In this approach,
the energy derivative can be obtained by the following equation:
\begin{eqnarray}
 \Delta E' = \langle \Phi_0 | (O_Ne^T)_c | \Phi_0 \rangle + \langle \Phi_0 | [\Lambda (O_Ne^T)_c]_c | \Phi_0 \rangle
 \label{energy_derivative}
\end{eqnarray}
where, $O_N$ is known as the derivative of normal ordered perturbed Hamiltonian with respect to external field of perturbation and
$\Lambda$ is an antisymmetrized de-excitation operator. This operator is given as 
\begin{eqnarray}
 \Lambda=\Lambda_1+\Lambda_2+ \dots+\Lambda_N=\sum_n^N \Lambda_n ,
\end{eqnarray}
with
\begin{eqnarray}
 \Lambda_m= \frac{1}{(m!)^2} \sum_{ij \dots ab \dots} \lambda_{ab \dots}^{ij \dots}{a_i^{\dagger}a_j^{\dagger} \dots a_b a_a} ,
\end{eqnarray}
where, $\lambda_{ab \dots}^{ij \dots}$ is the amplitude corresponding  to $\Lambda_m$.
In the CCSD framework, $\Lambda=\Lambda_1+\Lambda_2$. The explicit equations to solve the amplitudes of $\Lambda_1$
and $\Lambda_2$ are
\begin{eqnarray}
\langle \Phi_0 |[\Lambda (H_Ne^T)_c]_c | \Phi_{i}^{a} \rangle + \langle \Phi_0 | (H_Ne^T)_c | \Phi_{i}^{a} \rangle = 0,
\end{eqnarray}
\begin{eqnarray}
\langle \Phi_0 |[\Lambda (H_Ne^T)_c]_c | \Phi_{ij}^{ab} \rangle + \langle \Phi_0 | (H_Ne^T)_c | \Phi_{ij}^{ab} \rangle \nonumber \\
 + \langle \Phi_0 | (H_Ne^T)_c | \Phi_{i}^{a} \rangle \langle \Phi_{i}^{a} | \Lambda | \Phi_{ij}^{ab} \rangle = 0.
\label{lambda_2}
\end{eqnarray}
Once the amplitudes of $\Lambda$ are known, the desired property can be obtained from the Eq. (\ref{energy_derivative}).
\begin{table}[ht]
\caption{ Cutoffs for virtual spinors and basis sets used in our calculations.}
\begin{ruledtabular}
\newcommand{\mc}[2]{\multicolumn{#1}{#2}}
\begin{center}
\begin{tabular}{lccccr}
\mc{4}{c}{Basis} & \mc{2}{c}{Virtual}\\
\cline{1-4} \cline{5-6}  
Name & Nature & Cd & H & Cutoff (a.u.) & Spinors \\
\hline
A & DZ & dyall.ae2z & cc-pCVDZ & 500 & 157 \\
B & TZ & dyall.ae3z & cc-pCVTZ & 500 & 315 \\     
C & QZ & dyall.ae4z & cc-pCVQZ & 500 & 513 \\   
\end{tabular}
\end{center}
\end{ruledtabular}
\label{basis}
\end{table}

\section{Computational details}\label{comp}
We use a locally modified version of DIRAC10 \cite{dirac10} to solve the DHF equation and to generate the one- and two-electron integrals along with the necessary property integrals. 
The finite nucleus described by a Gaussian charge distribution is considered in our calculation \cite{visscher_1997}. 
The properties of interest are calculated using the $Z$-vector code developed in our group. We consider the bond length of CdH as 1.780 \AA \cite{length_ZnH}.
We have used the following basis sets: in the double-zeta: dyall.ae2z \cite{dyall_2010} for Cd, 
cc-pCVDZ \cite{dunning_1989} for H, in the triple-zeta (TZ) basis: dyall.ae3z \cite{dyall_2010} for Cd, and cc-pCVTZ \cite{dunning_1989} for H; 
in the quadruple-zeta (QZ) basis: dyall.ae4z for Cd, and cc-pCVQZ \cite{dunning_1989} basis for H. 
We correlate all the electrons and exclude the virtual spinors above a certain energy in the molecular calculations unless otherwise stated. 
The details of the basis sets used for CdH are given in Table \ref{basis}.

\begin{table}[ht]
\caption{Molecular-frame dipole moment, $\mu$ (in Debye) and the magnetic HFS constants (in MHz) of CdH. }
\begin{ruledtabular}
\newcommand{\mc}[1]{\multicolumn{#1}}
\begin{center}
\begin{tabular}{lccr}
Basis & $\mu$ & \mc{2}{c}{$^{111}$Cd} \\
\cline{3-4}  
 &  & A$_{\|}$ & A$_{\perp}$ \\
\hline
A & 0.61 & 4010 & 3595 \\
B & 0.73 & 4198 & 3762 \\
C & 0.76 & 4253 & 3817 \\
Expt. \cite{length_ZnH} &  & 4358(35) & 3966(3) \\
\end{tabular}
\end{center}
\end{ruledtabular}
\label{cdh_hfs}
\end{table}
\begin{table}[ht]
\caption{$\mathcal{P,T}$-odd interaction constants (W$_\mathrm{s}$ in kHz, E$_\mathrm{eff}$ in GV/cm, R in 10$^{18}$ /e.cm, and W$_\mathrm{M}$ in 10$^{33}$ Hz/e.cm$^2$ unit) of CdH.}
\begin{ruledtabular}
\begin{center}
\begin{tabular}{lccccr}
Basis & Nature & $W_\mathrm{s}$ & $E_\mathrm{eff}$ & R=$E_\mathrm{eff}$/W$_\mathrm{s}$ & $W_\mathrm{M}$\\
\hline
A & DZ & 11.3 & 10.3 & 220.4 & 0.76 \\
B & TZ & 13.3 & 11.9 & 216.3 & 0.81 \\
C & QZ & 14.0 & 12.2 & 210.7 & 0.82 \\
\end{tabular}
\end{center}
\end{ruledtabular}
\label{cdh_pt}
\end{table}

\begin{table*}[ht]
\caption{ The AIC properties of CdH at different cutoffs of virtual spinors (Basis: dyall.ae2z for Cd, cc-pCVDZ for H).}
\begin{ruledtabular}
\newcommand{\mc}[1]{\multicolumn{#1}}
\begin{center}
\begin{tabular}{lcccccr}
Virtual & \mc{2}{c}{Spinor} & A$_{\|}$ & $W_\mathrm{s}$ & $E_\mathrm{eff}$ & $W_\mathrm{M}$\\
\cline{2-3} 
Cutoff(a.u.) & Occupied & Virtual & (MHz) & (kHz) & (GV/cm) & (10$^{33}$ Hz/e.cm$^2$) \\
\hline
 50 & 49 & 121 & 3916 & 10.93 & 10.03 & 0.74 \\
 100 & 49 & 139 & 3976 & 11.13 & 10.21 & 0.75 \\
 200 & 49 & 145 & 3979 & 11.19 & 10.27 & 0.76 \\
 500 & 49 & 157 & 4010 & 11.25 & 10.31 & 0.76 \\
 1000 & 49 & 175 & 4029 & 11.31 & 10.36 & 0.76 \\
 No cutoff & 49 & 229 & 4049 & 11.40 & 10.46 & 0.77 \\
 No cutoff \footnote{using Dirac-Coulomb-Gaunt Hamiltonian} & 49 & 229 & 4050 & 11.43 & 10.24 &  0.77 \\ 
 \hline 
 50 & 19 & 121 & 3676 & 10.54 & 9.67 & 0.71 \\
 100 & 19 & 139 & 3681 & 10.55 & 9.68 & 0.71 \\
 200 & 19 & 145 & 3681 & 10.55 & 9.68 & 0.71 \\
 500 & 19 & 157 & 3682 & 10.56 & 9.69 & 0.71 \\
 1000 & 19 & 175 & 3682 & 10.56 & 9.69 & 0.71 \\
 No cutoff & 19 & 229 & 3682 & 10.56 & 9.69 & 0.71 \\
\end{tabular}
\end{center}
\end{ruledtabular}
\label{core_effect}
\end{table*}

\begin{table}
\caption{$\mathcal{P,T}$-odd properties of CdH as a function of bond length. (Basis used: dyall.ae2z for Cd and cc-pCVDZ for H, cutoff for virtual spinors = 500 a.u.)}
\begin{ruledtabular}
\begin{center}
\begin{tabular}{lccr}
Bond length & $E_\mathrm{eff}$ & $W_s$ & $W_M$ \\
(\AA) & (GV/cm) & (kHz) & (10$^{33}$ Hz/e.cm$^2$) \\ 
\hline
1.580 & 10.43 & 11.45 & 0.760 \\
1.680 & 10.43 & 11.41 & 0.766 \\
1.728 & 10.39 & 11.35 & 0.765 \\
1.780 ($r_e$) & 10.32 & 11.25 & 0.762 \\
1.834 & 10.21 & 11.13 & 0.756 \\
1.880 & 10.09 & 10.99 & 0.749 \\
1.980 & 9.76 & 10.61 & 0.726 \\
\end{tabular}
\end{center}
\end{ruledtabular}
\label{CdH_vib}
\end{table}

\begin{table}[ht]
\caption{Comparison of $\mathcal{P,T}$-odd interaction constants ($E_\text{eff}$ in GV/cm, $W_\text{s}$ in kHz and $W_\text{M}$ in 10$^{33}$ Hz/e.cm$^2$ unit) in ZnH, CdH and HgH. (Basis used: dyall.ae3z for Zn, Cd and Hg; 
    cc-pCVTZ for H. Cutoff for virtual spinors=500 a.u. Bond lengths for ZnH and HgH are 1.595 \AA \cite{length_ZnH} and  1.766 \AA \cite{length_ZnH}, respectively.)}
\begin{ruledtabular}
\newcommand{\mc}[3]{\multicolumn{#1}{#2}{#3}}
\begin{center}
\begin{tabular}{lcccccr}
Molecule  & \mc{2}{c}{$E_\mathrm{eff}$} & \mc{2}{c}{$W_\mathrm{s}$} & \mc{2}{c}{$W_\mathrm{M}$} \\
\cline{2-3} \cline{4-5} \cline{6-7}
× & DHF & Z-vector & DHF & Z-vector & DHF & Z-vector\\
\hline
ZnH & 1.7 & 2.13 & 1.4 & 1.83 & 0.21 & 0.27 \\
CdH & 9.5 & 11.91 & 10.3 & 13.29 & 0.65 & 0.81 \\
HgH & 106.8 & 123.37 & 241.2 & 284.34 & 2.94 & 3.21 \cite{talukdar_hgh} \\
\end{tabular}
\end{center}
\end{ruledtabular}
\label{comp_pt}
\end{table}

\section{Results and discussion}\label{res_dis}
We present the molecular dipole moment ($\mu$) and parallel and perpendicular components of the HFS constant of CdH and compare our results with available
experimental values \cite{length_ZnH} in Table \ref{cdh_hfs}. The magnitude of the dipole moment and the HFS constant increases as we move to a 
higher basis (i.e., from A to C). This is expected since the inclusion of
higher angular momentum basis functions can improve the configuration space. 
Our results of HFS constants are in good agreement with the available experimental values. However, 
the lowest deviation of the calculated HFS constants from the experimental values is obtained with the basis C (QZ, 500 a.u.). 
Our results of various $\mathcal{P,T}$-odd interaction parameters in CdH are presented in Table \ref{cdh_pt}. The most reliable values (calculated using basis C) of $E_\mathrm{eff}$, 
$W_\mathrm{s}$, $R$ and $W_\mathrm{M}$ are  12.2 GV/cm, 14.0 kHz, 210.7$\times$10$^{18}$ /e.cm and 0.82$\times$10$^{33}$ Hz/e.cm$^2$, respectively. 
The $\mathcal{P,T}$-odd molecular parameters in CdH are significantly large, which means that the
eEDM, S-PS nucleus-electron neutral current interaction and the interaction of the nuclear MQM with the magnetic field generated by electrons 
can contribute to the frequency shift in the $\mathcal{P,T}$-odd experiment. 
One can see from the electronic structure calculation that CdH can be a possible candidate for the experimental search of ``new physics'' in the lepton-sector of matter.
However, it may not be a choice for the MQM search since the isotopes of Cd having $I>1/2$ are very unstable.

We have mentioned earlier that the precise calculation of the AIC properties is not a trivial task due to the strong interelectronic correlations.
Therefore, it is important to investigate the systematic effects of the electron correlations and the virtual energy functions in the molecular calculations. 
To understand the correlation trend in the computed properties, we have performed two sets of calculations for CdH using the DZ basis 
(i.e., dyall.ae2z for Cd and cc-pCVDZ for H) at various cutoffs for the virtual 
spinors: firstly, correlating all the electrons and secondly, freezing the 1$s$-3$d$ electrons (i.e., correlating only 19 outer-electrons). 
We summarise these results in Table \ref{core_effect}. It is seen from this table that the magnitude of the A$_{\|}$, E$_\mathrm{eff}$, W$_s$, and W$_M$
increases with the number of virtual spinors in the given basis set for the all-electron correlation case. This is because, as the number of virtual
spinors (or the cutoff for virtual spinors) increases, the correlation space expands. In a larger correlation space, the electrons can be correlated more efficiently.
However, a different trend is observed in the frozen-core calculations. In this case, the computed properties slightly enhance as we increase the cutoff
of virtual spinors from 50 a.u. to 500 a.u., but further increase of the virtual cutoff does not enhance the magnitude of the properties anymore.
This means that the effect of high-energy virtual functions is more prominent when all the electrons are explicitly correlated in the molecular calculations.
From the Table \ref{core_effect}, it is also observed that the inner-core (1$s$-3$d$) electron correlations contribute significantly to the AIC properties
in the CdH molecule.

In the calculations of the AIC properties of CdH, we have not incorporated many important effects.
As a result, there could be some errors in our calculations.
The possible errors in our calculation may be caused by the following reasons:
(i) missing  of higher-order relativistic effects (the Breit/Gaunt interaction),
(ii) absence of higher-order correlation effects,
(iii) incompleteness of basis set, and
(iv) restriction of correlation space due to cutoff used for the virtual orbitals.
The $\mathcal{P,T}$-odd properties under study usually depend on the electron density
of the valence electron near the nuclear region and these properties are not very sensitive to the retardation and magnetic
effects \cite{quiney_1998_tlf, lindroth_1989}. 
However, we have calculated the mean-field Gaunt correction with DZ basis (see Table \ref{core_effect}) employing the DIRAC program package
which is found to be around 0.3\% and 2.2\% for W$_s$ and E$_\mathrm{eff}$, respectively, and negligible for W$_M$.
On the other hand, the error due to the absence of higher-order electron correlation effects can be
evaluated by comparing our values with the CCSD partial triples (CCSD(T)) or the full configuration interaction (FCI) results. But the CCSD(T) or
FCI calculation for CdH is too expensive to perform in the present work. However, in literature \cite{das_hgf, kudashov_2014},
this error was reported as around 3.5\% for some other but similar heavy diatomics. Therefore, we also expect a similar magnitude of the
error due to missing higher-order correlation effects in CdH. Similarly, another possible error yielded by basis set incompleteness 
can be assessed by comparing our results obtained using the A (DZ) and B (TZ) basis sets or the B (TZ) and C (QZ) basis sets. 
From Tables \ref{cdh_pt} and \ref{core_effect}, it is observed that
while going from the DZ to the TZ basis, the values of E$_\mathrm{eff}$, W$_s$, and W$_M$ are changed by 13.4\%, 15.4\%, and 6.2\%, respectively,
and while going from the TZ to the QZ basis, these values are changed by 2.5\%, 5.0\%, and 1.2\%, 
respectively. Thus, the error due to basis set incompleteness would not exceed 2.5\%, 5.0\%, and 1.2\% for our most reliable results 
of E$_\mathrm{eff}$, W$_s$, and W$_M$, respectively. It is also interesting to observe that the $\mathcal{P,T}$-odd S-PS nucleus-electron 
interaction parameter is more sensitive to the higher angular momentum basis functions than the E$_\mathrm{eff}$ and W$_M$. 
Furthermore, we have restricted the correlation space by excluding the virtual spinors with energy more than 
500 a.u. in the calculations of our most reliable results. It may yield some amount of error to our results. To decrease this type of error, we need to consider
the higher energy virtual spinors in our calculation which will be very much expensive and is beyond the scope of the present study.
However, we have performed the calculations for the $\mathcal{P,T}$-odd properties using the DZ basis (i.e., dyall.ae2z for Cd and cc-pCVDZ for H) at various cutoffs for the virtual 
spinors and summarised the results in Table \ref{core_effect} from which we can estimate this error. The high-lying virtual spinors with energy more than 500 a.u. contribute
1.3\% for S-PS interaction constant, 1.4\% for effective electric field and 1.3\% for MQM interaction constant.
In addition to the above-mentioned sources of error, the neglect of vibrational effects in the molecular calculation may also add some
amount of uncertainty to our results. The vibrational effects can be taken into account by doing a vibrational averaging of the calculated properties,
but it is beyond the scope of the present study.
As per our understanding, the vibrational effects may be important in a case when the molecular properties strongly depend 
on the internuclear distance of the molecule. In Table \ref{CdH_vib}, we summarise the results of the $\mathcal{P,T}$-odd properties of CdH at different internuclear
distances. From this table, we observe that for a change of around 3\% in the internuclear distance from the equilibrium bond length ($r_e$), 
the change in the values of the $\mathcal{P,T}$-odd
constants is within 1\%. This means that the studied properties of CdH do not have a strong dependence on the internuclear distance especially in the vicinity of $r_e$.
So, we expect that the vibrational correction to the calculated AIC properties of CdH would not be significant. Nevertheless, 
despite the possible cancellations of errors due to various effects,
we assess that the total uncertainty in our most reliable result is within 10\%. 


\begin{figure}[ht]
\centering
\begin{center}
\includegraphics[scale=.4, height=5cm]{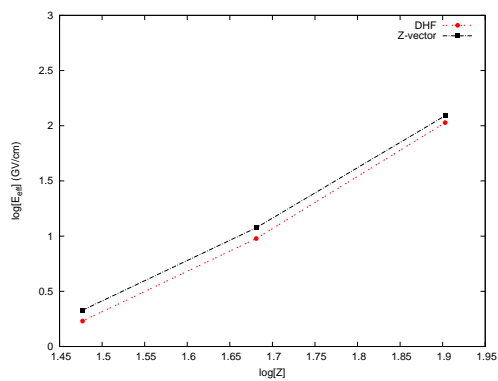}
\caption {Effective electric field, $E_\mathrm{eff}$ experienced by the unpaired electron in group-12 monohydrides.}
\label{CdH_Eeff}
\end{center}  
\end{figure}
\begin{figure}[ht]
\centering
\begin{center}
\includegraphics[scale=.4, height=5cm]{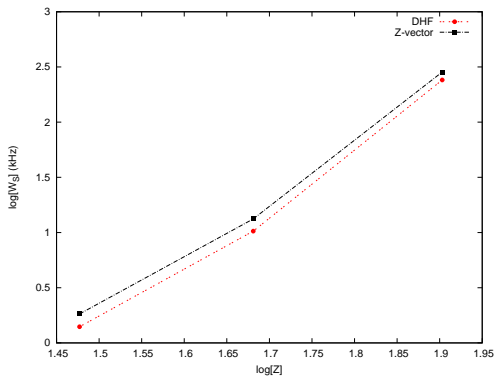}
\caption {S-PS nucleus-electron coupling parameter, $W_\mathrm{s}$ in group-12 monohydrides.}
\label{CdH_Ws}
\end{center}  
\end{figure}
\begin{figure}[ht]
\centering
\begin{center}
\includegraphics[scale=.4, height=5cm]{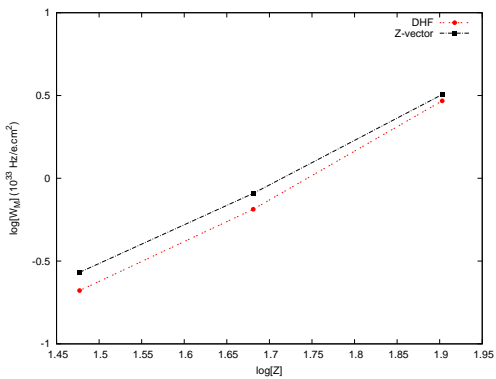}
\caption {Nuclear MQM-electron interaction parameter, $W_\mathrm{M}$ in group-12 monohydrides.}
\label{CdH_Wm}
\end{center}  
\end{figure}
 
We present the $\mathcal{P,T}$-odd interaction parameters of ZnH and HgH and compare them with those of CdH in Table \ref{comp_pt} to see the trend of the 
calculated properties in the monohydrides of group-12 elements. The $\mathcal{P,T}$-odd properties usually scale with nuclear 
charge Z of the heavy atom, and as shown in the Table \ref{comp_pt}, there is a monotonic increase in these properties 
from ZnH to HgH through CdH. We also plot log$_{10}$[$E_\text{eff}$], log$_{10}$[$W_\text{s}$], and log$_{10}$[$W_\text{M}$] 
against log$_{10}$[Z] in Figures \ref{CdH_Eeff}, \ref{CdH_Ws}, and \ref{CdH_Wm}, respectively. 
One can see the detailed discussion on the scaling of the $\mathcal{P,T}$-odd effects with Z given in Ref. \cite{berger_2019}.
We see from Table \ref{comp_pt} that the correlation contributions to $E_\text{eff}$, $W_\text{s}$ and $W_\text{M}$ in ZnH are around 20\%, 23\% and 22\%, respectively.
In CdH, electron correlation effects contribute around 20\%, 22\% and 20\% to $E_\text{eff}$, $W_\text{s}$ and $W_\text{M}$, respectively, whereas the said contributions 
are around 13\%, 15\% and 8\% to $E_\text{eff}$, $W_\text{s}$ and $W_\text{M}$, respectively in HgH. 
The DHF contribution to the total value of each molecular parameter of the 
group-12 monohydrides is significantly large in comparison to the electron-correlation contribution. One should also note that the said contribution in HgH is much higher
than that in ZnH and CdH. The weak screening effects of the 3$d$/4$d$/5$d$ electrons in these diatomic molecules can probably result in
a significantly large DHF contribution to the $\mathcal{P,T}$-odd interaction parameters \cite{sunaga_2018}.
Further, in Ref. \cite{sunaga_2017}, the reason for the high value of the $\mathcal{P,T}$-odd molecular parameter in HgH was
discussed using the Mulliken population analysis and the orbital interaction theory. The authors of the Ref. \cite{sunaga_2017} claimed that the large $s$-$p$ mixing in 
the singly occupied molecular orbital increases the effective electric field in a molecule. Thus, large $s$-$p$ mixing may be one of the reasons for the exceptionally high magnitude
of the $\mathcal{P,T}$-odd molecular parameters in HgH. The explicit study of the $\mathcal{P,T}$-odd effects in the HgH
molecule using the relativistic coupled-cluster method has already been done in Refs. \cite{sudip_hgh} and \cite{talukdar_hgh}. 
Although one would prefer HgH to CdH for the $\mathcal{P,T}$-odd frequency shift experiment due to the 
much higher values of the $\mathcal{P,T}$-odd constants in HgH than those in CdH, the possibility of CdH as a candidate for the 
same cannot be ruled out.


\par


\section{Conclusion}\label{conc}

We have studied the $\mathcal{PT}$ violating properties in the CdH molecule using the $Z$-vector method in the four-component relativistic coupled-cluster
framework and reported the corresponding molecular parameters. The value of $E_\text{eff}$, $W_\text{s}$ and $W_\text{M}$ in CdH reported by us are 12.2 GV/cm, 
14.0 kHz and $0.82\times$10$^{33}$Hz/e.cm$^2$, respectively, which are sufficiently large to be a possible candidate 
for the $\mathcal{P,T}$-odd experiment to reveal new physics beyond the standard model. We also compute the magnetic hyperfine structure 
constants of CdH and compare them with available experimental results to check the correctness of our calculations. 
Our reported HFS results are in good agreement with the corresponding experimental values. 
Our study shows that the correlation of the core-electrons is significantly important for the precise calculation of the 
AIC properties and the effect of the high-energy virtual spinors is more prominent in all-electron correlation treatment. Moreover,
the $\mathcal{P,T}$-odd interaction coefficients monotonically increase with the nuclear charge (Z) of the heavy atom in the monohydrides of group-12 elements.
\section*{Acknowledgement}
Authors acknowledge the resources of the Center of Excellence in Scientific Computing at CSIR-NCL. K.T. thanks the CSIR for 
the fellowship and Dr. Himadri Pathak and Dr. Sudip Sasmal for their insightful suggestion. N.V. acknowledges a grant from 
the DST.

\end{document}